\pdfoutput=1

\documentclass[aps,prb,reprint,showpacs,superscriptaddress]{revtex4-1}

\usepackage{graphicx}
\usepackage{graphics}
\usepackage{amsmath}
\usepackage{amssymb}
\usepackage{amsfonts}
\usepackage{dcolumn}
\usepackage{dsfont}
\usepackage{latexsym}
\usepackage{rotating}
\usepackage{color}
\usepackage{latexsym}
\usepackage{bbm}
\usepackage{subfigure}
\usepackage{float}
\usepackage{epsfig}
\usepackage{epsf}
\usepackage{psfrag}
\usepackage{bm}
\usepackage{amsthm}
\usepackage{eucal}
\usepackage{mathrsfs}
\usepackage{url}
\usepackage{braket}

\usepackage{color} 

\usepackage{hyperref}
\hypersetup{
colorlinks=true,final=true,
        linkcolor=blue,
        citecolor=blue,
        filecolor=blue,
        urlcolor=blue,
}
\begin{document}
\title{Nature of the spiral state, electric polarisation and magnetic transitions in Sr-doped YBaCuFeO$_5$: A first-principles study}
\author{Dibyendu Dey}
\email{dibyendu@phy.iitkgp.ernet.in}
\affiliation{Department of Physics, Indian Institute of Technology Kharagpur, Kharagpur - 721302, India}
\author{S. Nandy}
\affiliation{Department of Physics, Indian Institute of Technology Kharagpur, Kharagpur - 721302, India}
\author{T. Maitra}
\affiliation{Department of Physics, Indian Institute of Technology Roorkee, Roorkee - 247667, India}
\author{C. S. Yadav}
\affiliation{School of Basic Sciences, Indian Institute of Technology Mandi, Himachal Pradesh 175001, India}
\author{A. Taraphder}
\affiliation{Department of Physics, Indian Institute of Technology Kharagpur, Kharagpur - 721302, India}
\affiliation{Centre for Theoretical Studies and Centre for Nanoscience and Nanotechnology, Indian Institute of Technology Kharagpur, Kharagpur - 721302, India}

\date{\today}
\begin{abstract}

Contradictory results on the ferroelectric response of type II multiferroic YBaCuFeO$_{5}$, in its incommensurate phase, has of late, opened up a lively debate. There are ambiguous reports on the nature of the spiral magnetic state. Using first-principles DFT calculations for the parent compound within LSDA+U+SO approximation, the multiferroic response and the nature of spiral state is revealed. The helical spiral is found to be more stable below the transition temperature as spins prefer to lie in ab plane. The Dzyaloshinskii-Moriya (DM) interaction and the spin current mechanism were earlier invoked to account for the electric polarisation in this system. However, the DM interaction is found to be absent, spin current mechanism is not valid in the helical spiral state and there is no electric polarisation thereof. These results are in good agreement with the recent single-crystal data. We also investigate the magnetic transitions in YBa$_{1-x}$Sr$_x$CuFeO$_5$ for the entire range $0\le x\le 1$ of doping. The exchange interactions are estimated as a function of doping and a quantum Monte Carlo (QMC) calculation on an effective spin Hamiltonian shows that the paramagnetic to commensurate phase transition temperature increases with doping till $x=0.5$ and decreases beyond. Our observations are consistent with experimental findings.

\end{abstract}

\maketitle

\section{Introduction}

Type-II multiferroic materials, where ferroelectricity is induced by the spiral magnetic order ~\cite{katsu,mosto,khomski,tokura,kimura}, are expected to have various applications in magnetoelectric sensors, magnetic memory devices, etc. A prime objective is to design materials with high transition temperatures (T$_N$) for the spiral magnetic state. In systems which are not geometrically frustrated, the spiral state results from competing nearest-neighbour (J$_{NN}$) and next-nearest-neighbour exchange (J$_{NNN}$) interactions~\cite{herpin}. High values for such interactions are therefore desired for raising the T$_N$. However, in transition-metal oxides, such interactions are generally weak ~\cite{good} and T$_N$ for the spiral states are low ~\cite{kimura2,hur}. 

There are only two exceptions where magnetism-driven ferroelectricity is observed close to room temperature via a stable spiral state:  cupric oxide (CuO) \cite{kimura3,gio}, where a spiral phase is observed above 200 K, but only in a narrow window between 213 K to 230 K \cite{kimura3,forsyth}. The monoclinic symmetry with frustrating magnetic interactions and competing NN and NNN exchange interactions stabilize the magnetic spiral in this system around room temperature. The second one is the oxygen-deficient layered perovskite YBaCuFeO$_{5}$ (YBCFO), which displays multiferroism at unexpectedly high temperature \cite{kund,morin1}. Large electric polarisation \cite{kund,morin1} associated with the spiral phase has been reported in ceramic samples within a wide temperature range (almost ten times that of CuO) at zero magnetic field. However, there are contradictions regarding the direction of electric polarisation. Kundys et al. \cite{kund} proposed that the polarisation vector lies along the c axis due to the formation of dipole moments within the bipyramid, while Morin et al.\cite{morin1} suggest cycloidal nature of the spiral, tilted along c axis, favoring electric polarisation perpendicular to z direction. 

A very recent experiment on a single crystal \cite{chung} sample of parent YBCFO reach a completely different conclusion from what ceramic samples provided earlier. Their finding repudiates the ferroelectric nature of this compound as electric polarisation is entirely absent in the single crystal sample. However their experiments agree well with the previous magnetic measurements and confirm the presence of a spiral order. Subsequently, they suggested that the spins are rotated in the ab plane and form a magnetic spiral without any tilt along c. 

YBCFO crystallizes into a tetragonal structure with $P4mm$ symmetry\cite{morin2}. Experimentally it is not possible to assign any ordering of Fe/Cu ions due to their similar ionic radii. However, first-principles density functional theory calculations show that the bipyramidal layers are preferentially occupied by ferromagnetically coupled Fe-Cu pairs ~\cite{morin1} (Fig.~\ref{fig1}) within the experimentally observed commensurate magnetic structure. YBCFO undergoes two magnetic transitions \cite{kund,morin1,morin2,kawa,caig,mombru,ruiz}: higher temperature paramagnetic (PM) to commensurate (CM) antiferromagnetic phase transition occurs at $T_{N1}$ with the magnetic propagation vector \textbf{\textit{k$_{cm}$}}=$(1/2,1/2,1/2)$ \cite{morin1,morin2}. An incommensurate (ICM) magnetic  order sets in at a lower temperature $T_{N2}$. The neutron powder diffraction (NPD) data shows magnetic propagation vector \textbf{\textit{k$_{ icm}$}}=$(1/2, 1/2, 1/2 \pm q)$  in the ICM phase~\cite{morin1,chung,morin2}. However, ambiguity remains in the recent experiments \cite{morin1,kund,chung}; the connection between ferroelectric properties and the ICM phase and the nature of magnetic spiral in that regime are still open and unresolved issues. 

Morin et al.\cite{morin2} have recently observed that controlled introduction of chemical disorder into YBCFO during sample preparation, in the form of Fe-Fe or Cu-Cu impurity bonds, can enhance the stability of magnetic spiral state further, giving rise to a large increase of T$_{N2}$ beyond room temperature, as high as 310 K. Subsequently, Monte-Carlo simulations have shown that these impurity-bonds introduce a large out of plane antiferromagnetic exchange interaction which stabilizes the spiral state\cite{mucci1,mucci2}. Application of pressure or doping isovalent ions at the A sites are believed to have similar effect as chemical disorder described above. Enhancement of T$_{N2}$ by doping isovalent Sr at Ba sites has been observed in a very recent measurement~\cite{yadav}. 

From first-principles calculations we shed light on some of these issues, in particular, we address the following: (i) the spin orientation in the ab plane, (ii) the nature of extant magnetic order - helical spiral or cycloidal, (iii) whether Dzyaloshinskii-Moriya (DM) interaction is present and if there is a possible route to electric polarisation in this system. We compare our results with recent single-crystal data and find good agreement. We comment on earlier data on polycrystalline samples as well. 

We argue that the apparent discrepancy between different experimental data is likely to be an artifact of measurements on single-crystal versus polycrystal samples. Our theoretical results agree well with single crystal data. 
The rest of the paper is organized as follows. In Sec.~\ref{Method}, we discuss the methodology of our ab -\textit{initio} calculations in detail. In view of the recent disagreements on multiferroic nature of YBCFO, and the enhancement of transition temperature of the spiral phase with A site doping, we investigate the parent and doped compounds, using first-principles DFT calculations. The numerical results are discussed in Sec.~\ref{result}. First, we discuss the crystal and magnetic structure of both the parent and doped compounds in detail followed by the calculation of magnetic exchange interactions in CM phase to find the origin of the spiral phase. Next we study the effect of spin-orbit coupling (SOC) and calculate the anti-symmetric DM interaction parameter to identify the nature of the spin-spiral state in YBCFO. Finally, we study the effect of A site doping, followed by QMC calculation of an effective spin model whose exchange interactions are derived from DFT calculations. It would be useful to know how the exchange interactions evolve with doping and how they compete with each other in stabilizing or destabilizing the commensurate and incommensurate (spiral) magnetic phases. We also compare our results with existing experimental data. Finally in Sec.~\ref{summary}, we give a brief summary and outlook.

\section{Methodology}
\label{Method}
In order to study the electronic properties of the parent compound YBCFO and the Sr doped YBCFO (i.e. YBa$_{1-x}$Sr$_{x}$CuFeO$_{5}$), we have employed the first-principles density functional theory (DFT) calculations at various levels of approximations such as local spin-density approximation (LSDA), LSDA+U and LSDA+U+SO. For our DFT calculations we have used the plane-wave and psuedopotential based method as implemented in the Vienna ab-\textit{initio} Simulation Package (VASP) \cite{kresse} and projector augmented wave (PAW) potentials \cite{bloch}. The wave functions were expanded in the plane-wave basis with a kinetic-energy cutoff of 600 eV. Reciprocal space integration was carried out with a k-mesh of $8\times8\times4$. 

Electron correlation effects beyond LSDA, important to properly describe the ground state of transition metal oxides, are incorporated using LSDA+U \cite{duda} calculations where U is the on-site Coulomb correlation. We considered the value of U to be 5 eV for Fe and 8 eV for Cu~\cite{morin1}. Corresponding Hund’s coupling strengths (J) were set to J$_{Fe}$ =1 eV and J$_{Cu} $=0 for Fe and Cu respectively, as used in the previous literature~\cite{morin1}. Structural parameters were taken from experiments~\cite{yadav,pissas,kallias} for the parent and all the doped structures [Tab.~\ref{Tab1}]. Ionic positions for each structure are then optimized keeping the lattice constants at their respective experimental values.

We introduce SOC in our calculations to determine the spin easy-axis or easy-plane anisotropy of parent YBCFO within LSDA+U+SO approximation. In a previous report \cite{morin1}, it was suggested that DM interaction is the possible origin of electric polarisation in YBCFO. Since, DM interaction is associated with the SOC in the system, we have performed non-collinear DFT calculations with SOC to estimate the antisymmetric DM interaction parameter.

Further, we have carried out quantum Monte-Carlo (QMC) calculations using the loop algorithm of the ALPS 2.1 package ~\cite{ALPS} using the magnetic exchange interactions estimated from our DFT calculations to study temperature-dependence of magnetic susceptibility and specific heat of these systems. We also study the doping-dependence of PM to CM phase transition temperature. These calculations were performed on a $16\times16\times32$ supercell with periodic boundary condition.

\section{Results}
\label{result}
We have investigated the possible origins of the spin spirals in YBCFO and also predicted it's nature which supports recent experimental observations.

\subsection{Crystal and Magnetic Structure}
\begin{figure}
\centering
\includegraphics[width=9.5cm]{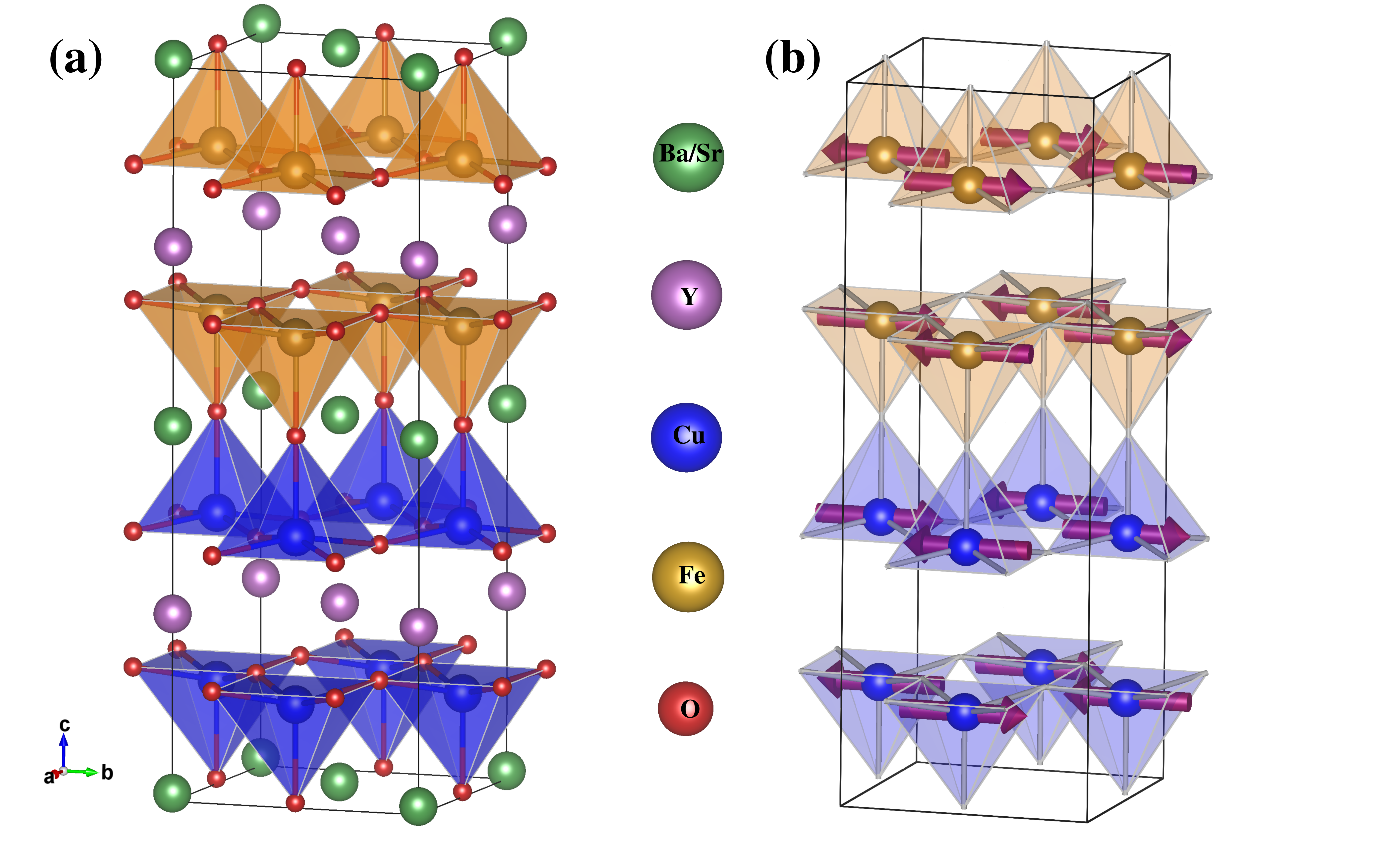}
\caption{(Color online) (a)$\sqrt{2}\times \sqrt{2}\times 2$ Supercell of YBa$_{1-x}$Sr$_{x}$CuFeO$_{5}$ showing FeO$_5$ and CuO$_5$ square pyramids in golden and blue colours respectively. (b) Experimentally observed magnetic structure of commensurate phase ~\cite{morin1,chung}.}
\label{fig1}
\end{figure}

YBa$_{1-x}$Sr$_{x}$CuFeO$_{5}$ has the layered perovskite structure with a noncentrosymmetric $P4mm$ space group \cite{morin1,yadav} for the entire range of doping as shown in Fig.~\ref{fig1}(a). In the tetragonal structure, two square pyramids FeO$_5$ and CuO$_5$ are connected via apical Oxygen forming a layer of bipyramids. Ba$ ^{2+} $/Sr$ ^{2+} $ ions go into the interstitial positions in between the two square pyramidal layers. These bipyramidal layers (or bilayers) are separated by a layer of Y$ ^{3+}$ ions.

In our ab-\textit{initio} calculations we have considered a supercell with a$_{s}$=$\sqrt{2}a $ and c$_{s}$=2c (a and c are the crystallographic unit cell parameters) containing four formula units, as depicted in Fig.~\ref{fig1}(a). This construction of the supercell was required to incorporate the CM magnetic structure shown in Fig.~\ref{fig1}(b) ~\cite{morin1}. In this magnetic phase, the bipyramidal units connected via apical Oxygen are preferentially occupied by Cu-Fe pairs and the interaction within these Cu-Fe dimers is ferromagnetic (FM).  Whereas Cu-Cu and Fe-Fe pairs separated by Y layers are antiferromagnetically coupled. Furthermore, all the nearest-neighbour (NN) interactions within the ab-plane are AFM in nature.

At lower temperatures (below T$_{N2}$) a spiral magnetic state is formed in which non-collinear spin arrangements within the bipyramidal layer (Cu-Fe pairs) appears along c direction ~\cite{morin1,morin2,chung}. However, inter-bipyramidal exchange (Cu-Cu and Fe-Fe) interactions remain AFM in the ICM phase. 
 
\begin{table}
\begin{center} 
\begin{tabular}{| p{2cm} || p{2cm} | p{2cm} | p{2cm} |} 
\hline 
x & a (\AA) & b (\AA) & c (\AA)\\ 
\hline\hline 
0.00 & 3.8711 & 3.8711 & 7.6629 \\ 
\hline
0.25 & 3.8603 & 3.8603 & 7.6595 \\
\hline
0.50 & 3.8506 & 3.8506 & 7.6517 \\
\hline
0.75 & 3.8408 & 3.8408 & 7.6342 \\
\hline
1.00 & 3.8317 & 3.8317 & 7.6076 \\ 
\hline 
\end{tabular}
\end{center} 
\caption{Experimental lattice constants of YBa$_{1-x}$Sr$_{x}$CuFeO$_{5}$ \cite{yadav,pissas,kallias}} 
\label{Tab1}
\end{table}

\subsection{Magnetic Exchange Interactions}
Consistent with the previous reports\cite{morin1,mucci1}, we also observe from our DFT total energy calculations that the parent compound YBCFO, in its commensurate magnetic phase, has the lowest energy when Fe$^{3+}$ and Cu$^{2+}$ ions are preferentially ordered in each bipyramid (Fig.~\ref{fig1}(a)). Hence we considered this structure for further calculations as discussed below. The nearest (J$_{NN}$) and next nearest neighbour (J$_{NNN}$) exchange interactions  were estimated by mapping the total energy difference between FM and AFM configurations for a pair of spins to the classical Heisenberg Hamiltonian\cite{morin1} which can be written as
\begin{eqnarray}
H = \frac{1}{2} \sum_{ij} J_{ij} (\mathbf{S}_{i} \cdot \mathbf{S}_{j})
\end{eqnarray}
Calculated J values are listed in Tab.~\ref{Tab2} for parent as well as for doped compounds. For the parent compound, the J-values are comparable to the previously reported values \cite{morin1}. Here $J_{p1}$ and $J_{p2}$ are the in-plane Cu-Cu and Fe-Fe NN exchange couplings respectively which are found to be AFM in nature. We also observe that the strongest exchange interaction is in the ab-plane ($J_{p1}$). The out-of-plane interactions along c ($J_{1}$, $J_{2}$, $J_{3}$ and $J_{4}$) are much weaker (almost one order of magnitude less) compared to the in-plane $J_{p1}$. 

To obtain insight into the origin of ICM phase, we studied the effect of next-nearest-neighbor (J$_{NNN}$) exchange interactions along the c direction on the magnetic ordering. One possible way to create magnetic frustration in the unfrustrated CM phase is to have a strong ferromagnetic J$_{NNN}$. However, our first-principles estimation of J$_{NNN}$ shows that this interaction (when ferromagnetic) is too small (Tab.~\ref{Tab2}) to cause enough frustration in destabilizing the CM phase in favour of an ICM one. The NNN interaction strengths for other structures with different Fe$^{3+}$/Cu$^{2+}$ ordering are also found to be either too weak or of wrong sign to stabilize a spiral spin state\cite{morin1}. In the following section we explore the other possible scenarios for the formation of a spiral state in this system.

\subsection{Effect of Spin-Orbit Coupling}

The directions of spin moments in the CM phase of YBCFO was a matter of debate as far as recent experimental observations are concerned ~\cite{morin1,chung}. In order to determine the spin easy-axis or easy-plane anisotropy of the CM phase in YBCFO, we consider spin-orbit (SO) coupling in our calculations within LSDA+U+SO approach. Comparing total energies for three configurations with spin moments directed along x, y and z directions in the CM phase, we observed that spin moments prefer to lie in the xy-plane rather than along z-direction. The energy difference ($E_{xy}-E_{z}$) between them is $\approx -3$ meV. In a very recent single crystal measurements on YBCFO, Chung et al.~\cite{chung} also observed the spin moments are aligned in the ab-plane which corroborates our theoretical findings.

Moving on to the crucial and highly debated issue concerning this material $-$ magnetic spiral state induced ferroelectricity in its ICM phase, we explore a plausible microscopic mechanism ~\cite{KNB,sergienko} which involves  antisymmetric DM interaction ~\cite{Dzya,moriya}. This interaction is a relativistic correction to the usual superexchange and its given by the following Hamiltonian
\begin{eqnarray}
H_{DM} = \sum_{ij} \mathbf{D}_{ij} \cdot [\mathbf{S}_{i} \times \mathbf{S}_{j}]
\end{eqnarray}
The strength of the interaction is proportional to the SOC constant. The DM interaction favours non-collinear spin ordering in perovskite manganites ~\cite{sergienko}. It also transforms the collinear state into a magnetic spiral in ferroelectric materials ~\cite{Zhao,Kadomtsev}. 

\begin{figure}
\centering
\includegraphics[width=9cm]{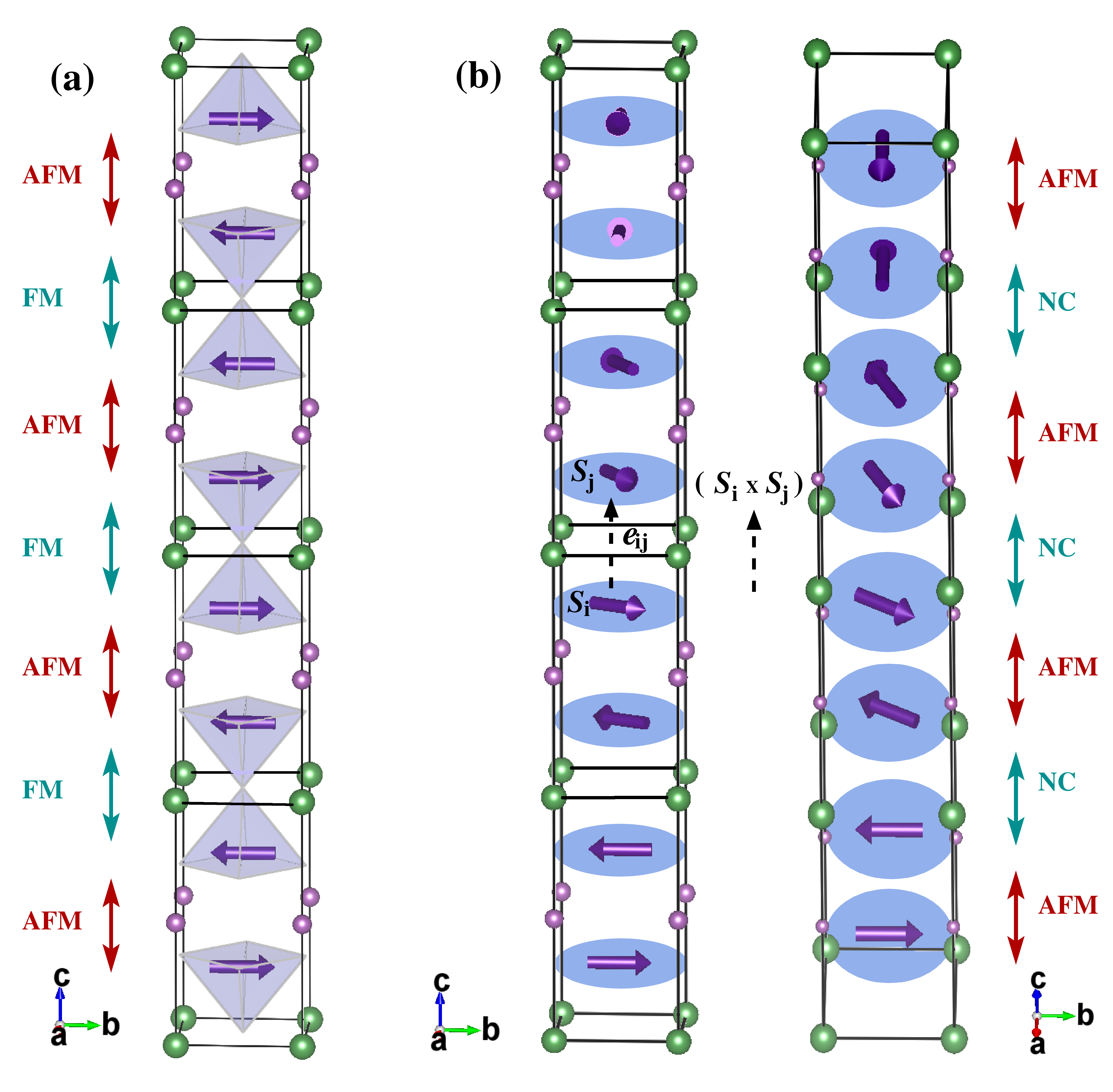}
\caption{(Color online) (a) Collinear magnetic order in CM phase (b) Helical magnetic spirals in ICM phase}
\label{fig2}
\end{figure}

We have performed DFT calculations with SOC to compute the antisymmetric DM interaction parameter (\textbf{D}) as described in the literature ~\cite{xiang}. The computed values for D$_x$, D$_y$ and D$_z$ are found to be negligibly small ($\approx 0.01$ meV) to cause any significant electric polarisation in YBCFO. This result rules out DM interaction as a possible origin of magnetically induced ferroelectricity in YBCFO as claimed in an earlier report ~\cite{morin1}. Since the spin easy axis remains in the ab-plane, spins will prefer to remain in the plane when the system transits from CM to ICM phase. This implies that the nature of spiral state would be helical rather than cycloidal in the ICM phase. The proposed helical spiral state has been depicted in Fig.~\ref{fig2}(b) where the non-colinearity appears within the ferromagnetically coupled bipyramids.

Based on the cycliodal spiral state a spin-current model\cite{KNB} was proposed earlier\cite{morin1} to explain the existence of electric polarisation in the ICM phase of YBCFO. According to this mechanism, the polarisation direction is given by $\mathbf{e}_{ij} \times (\mathbf{S}_{i} \times \mathbf{S}_{j})$, where $\mathbf{e}_{ij}$ is the vector connecting the i and j sites (see Fig.~\ref{fig2}(b)) and spin indices are given by $\mathbf{S}_{i}$ and $\mathbf{S}_{j}$ respectively. However, if the spiral state is helical as proposed here (shown in Fig.~\ref{fig2}(b)), one can immediately see that the direction of $(\mathbf{S}_{i} \times \mathbf{S}_{j})$ is along c direction and hence parallel to $\mathbf{e}_{ij}$ which will give rise to zero electric polarisation.  Therefore, in both cases the polarisation is expected to be absent. From the above discussions we therefore conclude that (i) the spins are oriented in the ab plane, (ii) spin spiral state in the ICM phase of YBCFO is helical rather than cycloidal~\cite{morin1,morin2}, (iii) there is no electric polarisation in the ICM phase of YBCFO. These conclusions agree very well with the recent single crystal measurements~\cite{chung}.

\subsection{Effect of Sr doping on Magnetic Exchange Interaction}
In order to see the effect of Sr doping on the various magnetic exchange interactions present in YBCFO, we have performed DFT calculations of YBa$_{1-x}$Sr$_{x}$CuFeO$_{5}$ compound for $x=0,0.25,0.5,0.75,1$. We have calculated the magnetic exchange interactions for the doped compounds considering the CM magnetic structure as shown in Fig.~\ref{fig1}(b) within LSDA+U (see Tab.~\ref{Tab2}). For $x=0.5$ case we observed that alternate layers containing only Sr or Ba has lower energy than both layers having equal number Sr and Ba ions. So we have considered the former in the $x=0.5$ structure. From Tab.~\ref{Tab2} we see that $J_{p1}$ increases with doping up to $x=0.5$ and then decreases. One would therefore expect the PM to CM phase transition temperature ($T_{N1}$) to also follow a similar trend. Along $c$, two types of NN interactions exist in the CM magnetic phase; one is the intra-bipyramidal ferromagnetic exchange interaction between Fe and Cu and the other is inter-bipyramidal antiferromagnetic exchange interaction between Fe-Fe and Cu-Cu pairs. We discuss below how various exchange interactions along c evolve with respect to doping.

\begin{table}
\begin{center} 
\begin{tabular}{|p{1cm}|p{1cm}|p{1cm}|p{1cm}|p{1cm}|p{1cm}|p{1cm}|p{1cm}|} 
\hline 
x & J$ _{1} $ & J$ _{2}$ & J$ _{3} $ & J$ _{4} $ & J$ _{p1} $ & J$ _{p2} $ & J$ _{NNN} $ \\ 
\hline\hline 
0.00 & -2.016 & -2.016 & 14.01 & 2.86 & 149.5 & 8.78 & -0.069 \\ 
0.25 & -2.383 & -1.783 & 13.52 & 2.68 & 150.5 & 8.95 & -0.087 \\
0.50 & -2.766 & -1.646 & 13.10 & 2.59 & 155.5 & 9.15 & -0.117 \\
0.75 & -2.511 & -1.940 & 12.59 & 2.40 & 146.7 & 9.27 & -0.095 \\
1.00 & -2.325 & -2.325 & 12.27 & 2.31 & 147.6 & 9.41 & -0.087 \\ 
\hline 
\end{tabular}
\end{center}
\caption{NN and NNN exchange interaction strengths in YBa$_{1-x}$Sr$_{x}$CuFeO$_{5}$ in meV.} 
\label{Tab2}
\end{table}

\subsubsection{NN-FM interaction}
\begin{figure}
\centering
\includegraphics[width=9.0cm]{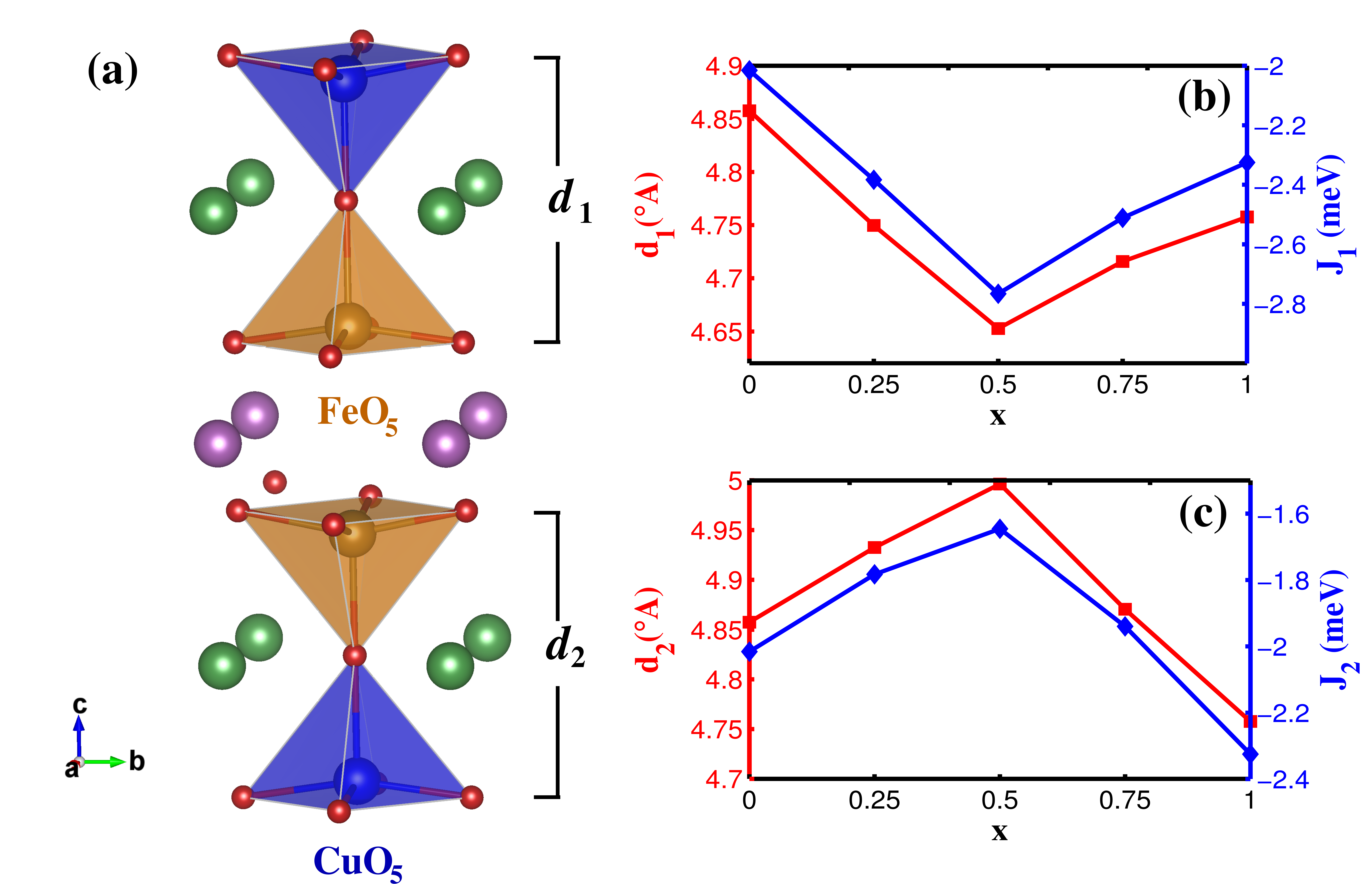}
\caption{(Color online) (a) d$_1$ (d$_2$) marked in the crystal structure as the thickness of the Sr (Ba)-containing bipyramidal layers (b) Variation of d$_1$ and $J_1$ with doping (c) d$_2$ and $J_2$  with doping.}
\label{fig3}
\end{figure}

As we replace Ba with Sr having smaller ionic size, the thickness of bipyramidal layer starts to shrink from the corresponding value in the parent compound YBCFO. For example, at x=0.25 the thickness of the bilayer containing Sr (d$_{1}$) decreases, whereas the thickness of the bilayer not containing Sr (d$_{2}$) increases (see Fig.~\ref{fig3}). This change in the thickness is expected to affect the corresponding exchange interactions. As we see in Fig.~\ref{fig3}(b), ferromagnetic exchange interaction strength between Fe-Cu pairs ($J_1$) corresponding to the bilayer containing Sr increases whereas the same for the bilayer not containing Sr ($J_2$) decreases (Fig.~\ref{fig3}(c)). This trend is followed for x=0.5 case as well where Sr and Ba segregate into different bilayers (see Fig.~\ref{fig3}(b)). 

For $x>0.5$, when both the bilayers contain Sr ions, the dependence of (d$_{1}$) and (d$_{2}$) on doping is opposite to what is observed in $x\le 0.5$ case. So are the behaviour of $J_1$ and $J_2$. Note that at x=1.0 (i.e. YSrCuFeO$_{5}$) the FM exchange of both bilayers are equal and is slightly more than the  corresponding exchange interactions for x=0 (i.e. YBaCuFeO$_{5}$). As it has been reported in the literature~\cite{morin1,morin2} that the ICM phase is connected with the appearance of non-collinearity within the bilayers, the enhancement of FM exchange in YSrCuFeO$_{5}$ (YSCFO) could imply that the CM phase would be more stable in this compound. The experimental reports on YSCFO compound \cite{pissas,kallias} indeed show that there is no CM to ICM phase transition.

\subsubsection{NN-AFM interaction}
\begin{figure}
\centering
\includegraphics[width=9.0cm]{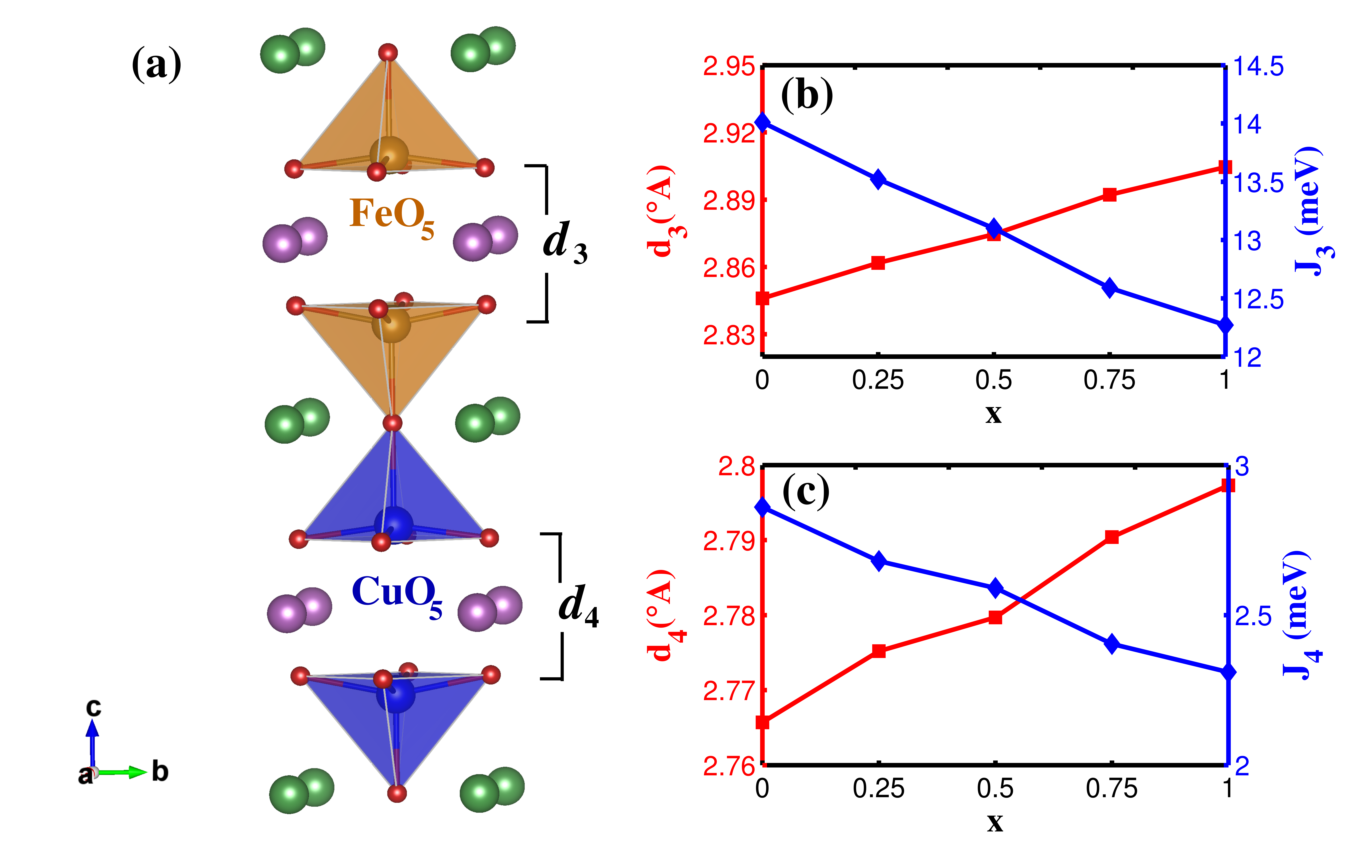}
\caption{(Color online) (a) d$_3$(d$_4$) marked in the crystal structure as the inter bilayer distance separated by a layer of Y-ions (b) Variation of d$_3$ and $J_3$ with doping, (c) d$_4$ and $J_4$ with doping.}
\label{fig4}
\end{figure}

In addition to its effect on the Fe-Cu ferromagnetic exchange interactions within the bipyramidal layers, Sr doping is also seen to influence the AFM exchange interactions present between Cu-Cu and Fe-Fe ions in the adjacent bilayers as shown in Fig.~\ref{fig4}. The inter-bipyramidal distances (d$_{3}$ and d$_{4}$) are observed to increase with doping almost linearly. The corresponding AFM exchanges (J$_{3}$ and J$_{4}$) decrease monotonically with Sr doping. This effect is similar to what is observed in a recent experimental measurement in YBCFO with Fe/Cu chemical disorder \cite{morin2}. The authors have reported that on increasing disorder the inter-bilayer distance increases whereas intra-bilayer distance decreases.

\begin{figure}
\centering
\includegraphics[width=7.5cm]{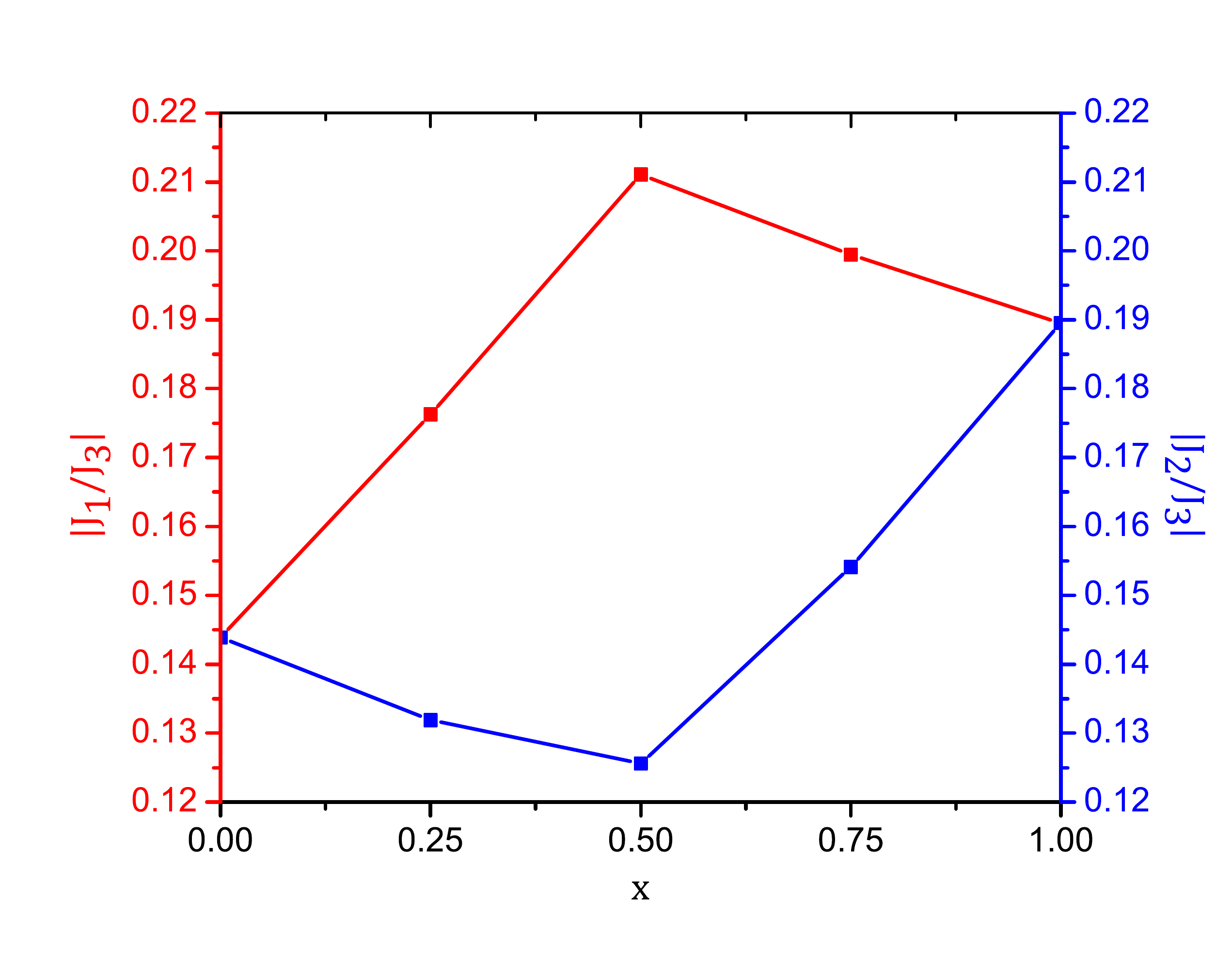}
\caption{(Color online) Variation of $J_1$/$J_3$  and $J_2$/$J_3$  with doping.}
\label{fig5}
\end{figure}

In Fig.~\ref{fig5} we have plotted the ratios $J_1$/$J_3$ and $J_2$/$J_3$ with doping, and we observe that these two ratios follow opposite trend with respect to the doping concentration. This indicates that the competition between these two interaction ratios may decide the magnetic phase diagram of these systems which is more complex than that due to Fe/Cu chemical disorder.

\subsubsection{NNN interaction}
\begin{figure}
\centering
\includegraphics[width=9.0cm]{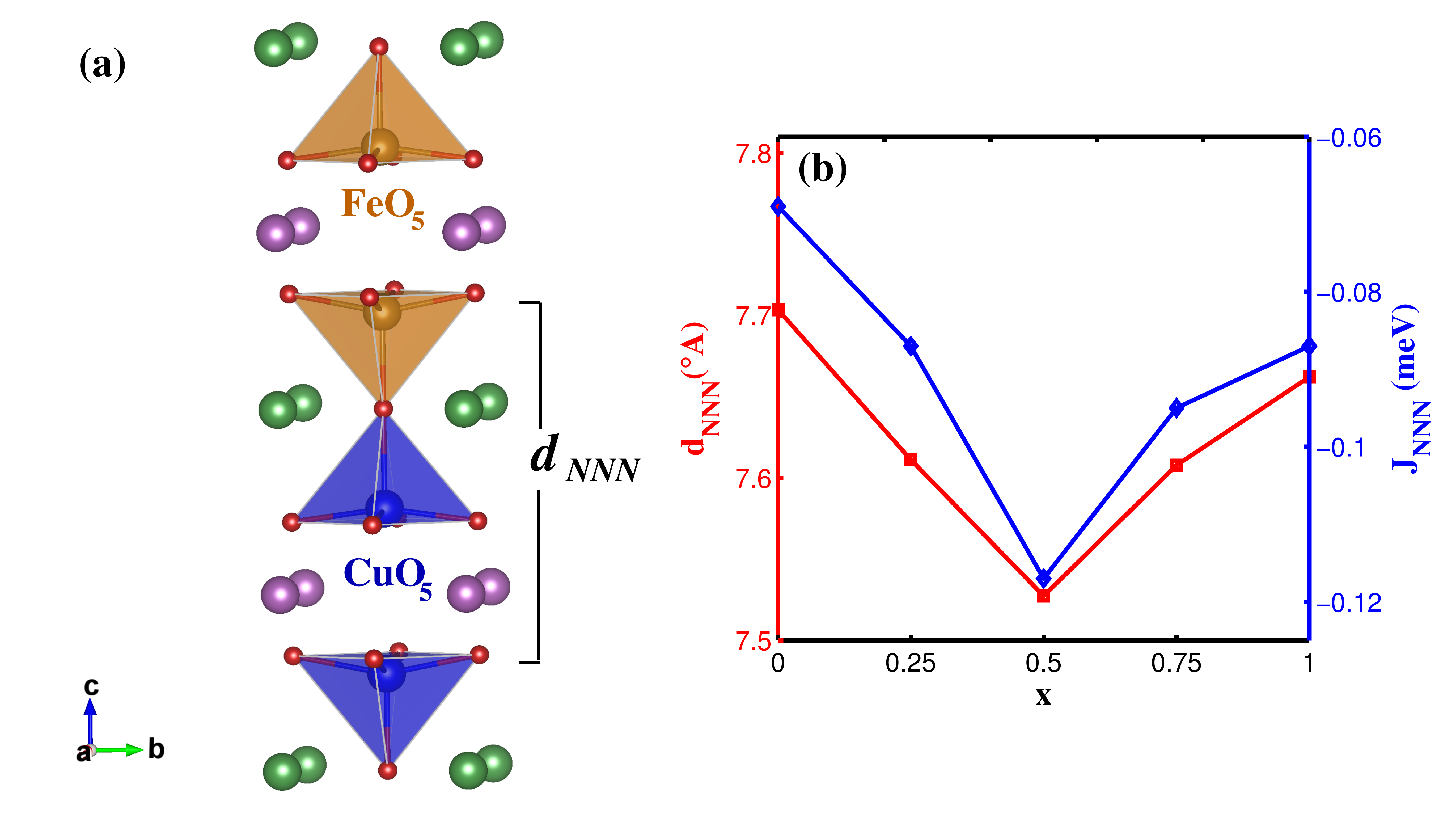}
\caption{(Color online) (a) d$_{NNN}$ shown in the crystal structure (b) Variation of $J_{NNN}$ with doping.}
\label{fig6}
\end{figure}

Finally, it is important to study the next nearest neighbour (NNN) exchange interaction with doping as it can play a very important role in stabilizing the ICM phase. In Fig.~\ref{fig6} we plot the calculated NNN exchange interaction strengths ($J_{NNN}$) with respect to doping along with the corresponding NNN distance (d$_{NNN}$). We observe that the interaction is ferromagnetic throughout the entire range of doping which is required to generate frustration in the CM phase. Though the strength by far is the smallest and it is seen to increase with Sr doping up to x=0.5 and then decreases with higher doping. This result implies that the spiral state in the parent compound may get stabilized with doping at least up to x=0.5 with Sr doping.   
\subsection{Finite Temperature Calculations: QMC results}
\begin{figure*}
\centering
\includegraphics[width=15cm]{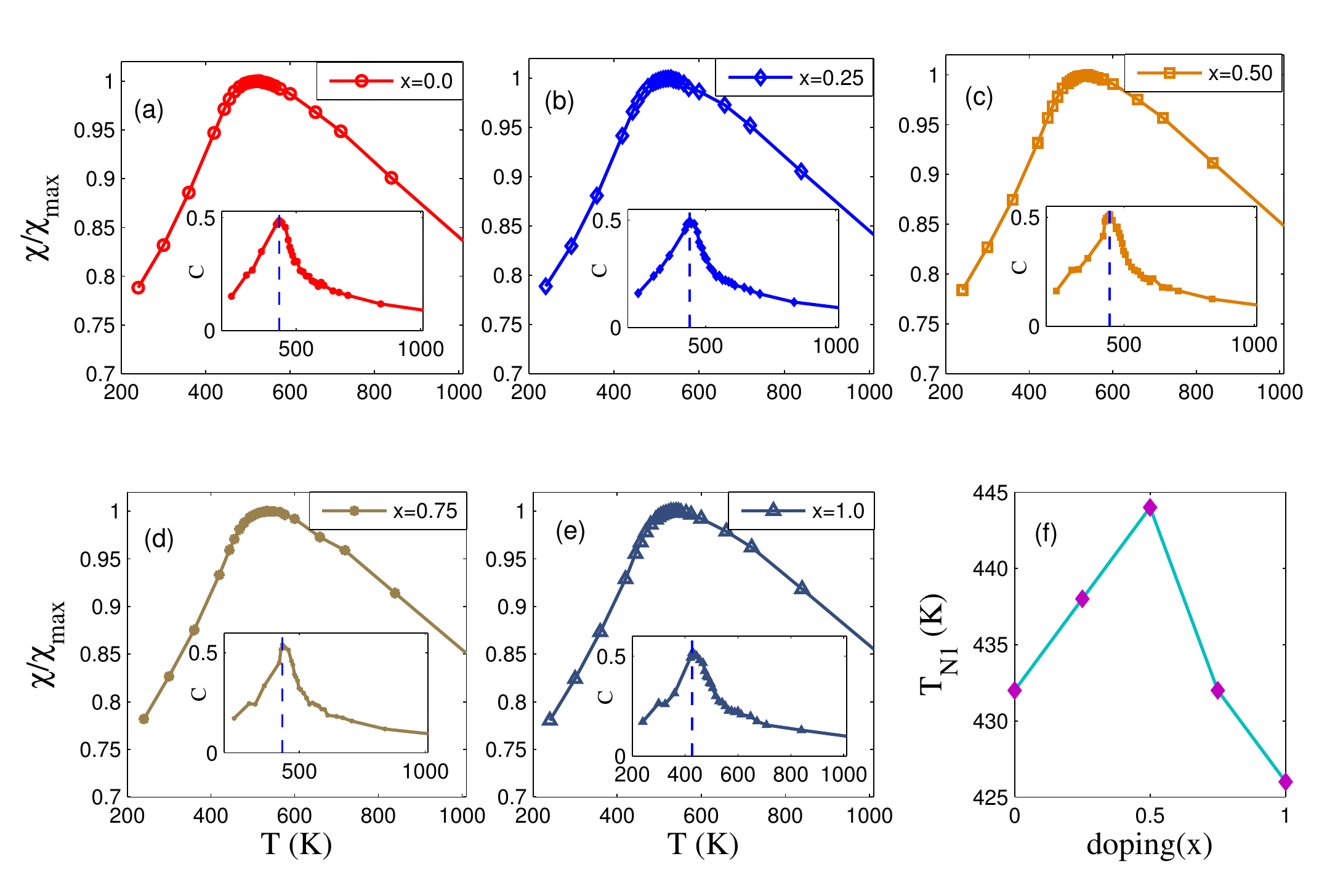}
\caption{(Color online) Magnetic susceptibility as a function of temperature at Sr doping (a) x=0.00, (b) x=0.25, (c) x=0.50 (d) x=0.75 (e) x=1.00. Variation in the specific heat with temperature is shown in the inset, (f) Variation of T$_{N1}$ (from the peak of the specific heat, dashed vertical line in (a)-(e) insets) with $x$.}
\label{fig7}
\end{figure*}

At last, we have performed QMC calculations with a Heisenberg spin Hamiltonian using the magnetic exchange interactions estimated from our DFT calculations as discussed above. In Fig.~\ref{fig7}(a)-(e) we present the temperature-dependence of the magnetic susceptibility and specific heat (inset) at different Sr-doping. We identify the magnetic transition from the peak in the susceptibility and specific heat curves. In Fig.~\ref{fig7}(f) we plot the variation of T${_{N1}}$ with doping. We observe that the transition temperature increases with doping up to $x=0.5$ and reaches its maximum at 444 K. However, on further doping, the transition temperature starts to decrease. A comparison of calculated T$_{N1}$s for the parent compounds (YBCFO and YSCFO) match very well with the previous experimental and theoretical results~\cite{morin1,morin2,pissas} on them. Experimental observations on PM to CM phase transition for YBCFO \cite{morin1} and YSCFO \cite{pissas} show that the T${_{N1}}$ is higher in case of YBCFO than that for YSCFO. This is also borne out quite well from our QMC results. 

We would like to note that, J$_{NNN}$ has also been included in our QMC calculation but it fails to create enough frustration for stabilizing spiral phase.

\section{Conclusions}
\label{summary}
Using first-principles density functional theory, we have studied in detail the nature of spin spiral state in the parent YBCFO and its correlation to the ferroelectric behaviour of the compound. We have clearly identified the magnetic spiral state to be a helical one where the spin moments rotate in the ab-planes without any out of plane components. We also rule out the role of DM interaction in inducing ferroelectricity in the compound as the interaction strength is found to be exceedingly small. We also show that the proposed mechanism based on spin current model\cite{KNB} can not lead to finite electric polarisation in a helical spiral spin state. These findings are corroborated by the very recent experimental observations on a single crystal ~\cite{chung} YBCFO sample, where the authors did not find any electric polarisation in the ICM phase and the spiral state is found to be helical rather than cycloidal. 

In another recent experimental observation, it has been reported that the stability of the spiral state can be tuned with A site doping ~\cite{yadav}. A systematic study of magnetic phase transitions in Sr-doped YBCFO for the entire range of doping has therefore been performed by us using DFT and QMC calculations. The structural changes due to doping is seen to affect various magnetic exchange interactions and the corresponding transition temperatures. We have evaluated various NN and NNN exchange interactions between Fe-Cu, Fe-Fe and Cu-Cu spins, for the experimentally observed CM magnetic structure. We find a strong dependence of the exchange interactions on Sr doping. The Fe-Cu exchange (J$_1$) along c direction, within the bipyramidal layer containing Sr, is ferromagnetic. It is found to increase with doping up to  $x \leq 0.5$. However, the magnetic exchange (J$_2$), within a bipyramidal layer containing Ba, is seen to decrease. The reverse trend is observed for both J$_1$ and J$_2$ in the region $x > 0.5$. In addition, the inter-bipyramidal (separated by Y) AFM exchange couplings J$_3$ and J$_4$, between Cu-Cu and Fe-Fe pairs respectively, decrease monotonically with Sr doping. 

Quite importantly, J$_{NNN}$ is ferromagnetic; its value is found to increase with doping up to $x=0.5$ though the strength is small compared to NN exchanges. It's enhancement with doping, therefore, indicates that the spiral state in the parent compound might be more stable with Sr doping at Ba sites atleast upto $x=0.5$. From a QMC calculation on a spin Hamiltonian, whose spin exchange parameters are derived from DFT, we observe that the PM to CM phase transition temperature T$_{N1}$ increases with doping up to $x=0.5$ and decreases beyond. These observations are consistent with experimental data~\cite{morin1,yadav,pissas,kallias} and show the way forward in doping-control of transition temperature in YBCFO.

\section{Acknowledgment}
DD, SN, and AT acknowledge the computing facility from DST-Fund for Improvement of S \& T infrastructure (phase-II) in the Department of Physics, IIT Kharagpur. DD acknowledges DST, India for INSPIRE research fellowship. SN acknowledges MHRD, India for research support.  AT acknowledges CSIR  (India)  for funding through the project grant no: 03 (1373)/16/EMR-II, Dt: 10-05-2016.


\begin{thebibliography}{99}

\bibitem{katsu} H. Katsura, N. Nagaosa, and A. V. Balatsky, Phys. Rev. Lett. {\bf 95}, 057205 (2005).
\bibitem{mosto} M. Mostovoy, Phys. Rev. Lett. {\bf 96}, 067601 (2006).
\bibitem{khomski} D. Khomskii, Physics {\bf 2}, 20 (2009).
\bibitem{tokura} Y. Tokura and S. Seki, Advanced Materials {\bf 22}, 1554 (2010).
\bibitem{kimura} T. Kimura, Annu. Rev. Mater. Res. {\bf 37}, 387 (2007).
\bibitem{herpin} A. Herpin, Theorie du magnetisme (Presses Universitaires de France, Paris, 1968).
\bibitem{good} J. B. Goodenough, Magnetism and the Chemical Bond (Wiley, Cambridge, MA, 1963).
\bibitem{kimura2} T. Kimura, T. Goto, H. Shintani, K. Ishizaka, T. Arima and Y. Tokura, Nature {\bf 426}, 55 (2003)
\bibitem{hur} N. Hur, S. Park, P. A. Sharma, J. S. Ahn, S. Guha and S-W. Cheong, Nature {\bf 429}, 392 (2004)
\bibitem{kimura3} T. Kimura, Y. Sekio, H. Nakamura, T. Siegrist, and A. Ramirez, Nat. Mater. {\bf 7}, 291 (2008).
\bibitem{gio} G. Giovannetti, S. Kumar, A. Stroppa, J. van den Brink, S. Picozzi, and J. Lorenzana, Phys.Rev.Lett. {\bf 106}, 026401 (2011).
\bibitem{forsyth} J. B. Forsyth, P. J. Brown, and B. M. Wanklyn, J. Phys. C: Solid State Phys. {\bf 21}, 2917 (1988).
\bibitem{kund} B. Kundys, A. Maignan, and C. Simon, Appl. Phys. Lett. {\bf 94}, 072506 (2009).
\bibitem{morin1} M. Morin, A. Scaramucci, M. Bartkowiak, E. Pomjakushina, G. Deng, D. Sheptyakov, L. Keller, Rodriguez-Carvajal, N. A. Spaldin, M. Kenzelmann, K. Conder, and M. Medarde, Phys. Rev. B {\bf 91}, 064408 (2015).
\bibitem{chung} Yen-Chung Lai et al. J. Phys.: Condens. Matter {\bf 29}, 145801 (2017).
\bibitem{morin2} M. Morin, E. Canevet, A. Raynaud, M. Bartkowiak, D. Sheptyakov, B. Voraksmy, M. Kenzelmann, E. Pom-jakushina, K. Conder, and M. Medarde, Nat. Commun. {\bf 7}, 13758 (2016).
\bibitem{kawa} Y. Kawamura, et al., J. Phys. Soc. Jpn. {\bf 79}, 073705 (2010).
\bibitem{caig} V. Caignaert, et al.  J. Solid State Chem. {\bf 114}, 24 (1995).
\bibitem{mombru} A.W. Mombru, et al.  J. Phys.: Condens. Matter {\bf 10}, 1247 (1998).
\bibitem{ruiz} M. J. Ruiz-Aragon, U. Amador, J. L. Martinez, N. H. Andersen, and Ehrenberg,  Phys. Rev. B {\bf 58}, 6291 (1998).
\bibitem{mucci1} A. Scaramucci et al., arXiv:1610.00783
\bibitem{mucci2} A. Scaramucci et al., arXiv:1610.00784
\bibitem{yadav} Surender Lal, Sanjay K. Upadhyaya, K. Mukherjee and C.S.Yadav, EPL {\bf 117}, 67006 (2017).
\bibitem{kohn} P. Hohenberg and W. Kohn, Phys. Rev. {\bf 136}, B864 (1964)
\bibitem{kresse} Kresse and J. Hafner, Phys. Rev. B {\bf 47}, 558 (1993); G. Kresse
and J. Furthmüller, Comput. Mater. Sci. {\bf 6}, 15 (1996); G. Kresse
and J. Furthmüller, Phys. Rev. B {\bf 54}, 11169 (1996)
\bibitem{bloch} P. E. Blöchl, Phys. Rev. B {\bf 50}, 17953 (1994); G. Kresse and D. Joubert, Phys. Rev. B {\bf 59}, 1758 (1999).
\bibitem{duda} S. L. Dudarev, G. A. Botton, S. Y. Savrasov, C. J. Humphreys and A. P. Sutton, Phys. Rev. B {\bf 57}, 1505 (1998).
\bibitem{pissas} M Pissas, G Kallias, A Simopoulos, D Niarchos, E Devlin and R Sonntag, J. Phys.: Condens. Matter {\bf 10}, 10317 (1998).
\bibitem{kallias} G. Kallias et al. Physica B {\bf 234-236} 608-610 (1997).
\bibitem{ALPS} Bauer, B.et al. The alps project release 2.0: open source
software for strongly correlated systems. Journal of Statistical Mechanics: Theory and Experiment 2011, P05001 (2011)
\bibitem{KNB} Hosho Katsura, Naoto Nagaosa, and Alexander V. Balatsky, Phys. Rev. Lett. {\bf 95}, 057205 (2005)
\bibitem{sergienko} I. A. Sergienko and E. Dagotto, Phys. Rev. B {\bf 73}, 094434 (2006).
\bibitem{Dzya} I. Dzyaloshinskii, Sov. Phys. JETP {\bf 19}, 960 (1964).
\bibitem{moriya} T. Moriya, Phys. Rev. {\bf 120}, 91 (1960)
\bibitem{Zhao} T. Zhao et al., Nature Materials {\bf 5}, 823 (2006).
\bibitem{Kadomtsev} A.M. Kadomtseva et al., JETP Lett. {\bf 79}, 571 (2004)
\bibitem{xiang} H. J. Xiang et al., Phys. Rev. B {\bf 84}, 224429 (2011)
\end{thebibliography}
\end{document}